# A NEW CLASS OF ORBITING VERY LARGE ULTRA-LIGHT ORIENTABLE MW ANTENNAS IN LOW EARTH ORBITS

[1]Giovanni Perona, [2]Angelo Raffaele Meo, [3]Marco Allegretti, [4]Isabella Bordi, Elena Marengo, Bruno Mazzetti, [1]Mario Scovazzi

[1]*CSP innovazione nelle ICT, Strada del Lionetto, 6, Torino, Italia*

[2]*Politecnico di Torino, Department of Control and Computer Engineering, C.so Duca degli Abruzzi, 24, Torino, Italia*

[3]*Politecnico di Torino, Electronics Dept. C.so Duca degli Abruzzi, 24, Torino, Italia*

[4]*Envisens Tecnologies S.r.l., Corso Ciro Menotti, 4, Torino, Italia*

(e-mail: giovanni.perona@formerfaculty.polito.it, meo@polito.it, marco.allegretti@polito.it, is.bordi@gmail.com, elena.marengo68@gmail.com, bmazzetti49@gmail.com, mario.scovazzi@csp.it)

## ABSTRACT

The purpose of the following proposal is to create a new class of antennas of low weight but with a large transmitting and receiving surface made of metallized fabric.

This class of antennas will be stationed in low satellite orbits, at about 2000 km from sea level and will be easily steerable by exploiting the interaction between the earth's magnetic field and an electromagnetic field generated on the antenna.

The possibility of orienting very large ultra-light balloon-like MW antennas, orbiting in the Earth magnetic field, through electric currents flowing in conductive wires embedded in the antennas walls, are examined. It is shown that such currents may easily provide the needed forces without the necessity of mechanically orienting the antenna itself.

The antenna is extremely light, inexpensive, easily directional.

The antenna has the shape of a spheroid in which the southern hemisphere has a parabolic shape made of metallized fabric; the northern hemisphere, transparent to the electromagnetic radiation, has a radius of curvature such that the north pole coincides with the focus of the parabola.

## INTRODUCTION

In the '80, ESA promoted a study concerning the deployment of a 15 m diameter antenna, using the inflatable Space Rigidized technology. The reflecting antenna was supposed to be attached to a radio astronomy satellite named QUASAT, orbiting the earth on an elliptical orbit and to be used with the VLBI network in Europe and the US. The satellite was intended to operate at different wavelength: 1.35 cm, 6 cm, 18 cm and 92 cm [1].

More recently, a scientific radio telescope satellite, named Spectr-R, with a 10 m antenna, has been launched in a highly elliptical orbit on 18 July 2011. However the antenna was a "traditional" antenna consisting of a solid central mirror of 3 m diameter surrounded by 27 solid petals made of carbon fiber [2].

Following previous studies, JPL in the '90 and at the beginning of this century produced a large number of documents concerning inflatable structures. In particular in December 1998 [3], the Arise team prepared a very detailed study concerning the possibility of putting in orbit a satellite named ARISE. Such a satellite was supposed to have a 30 m antenna and to be placed in an elliptical orbit with a 5000 km perigee and a 40000 km apogee. Other studies on inflatable structures have been produced in more recent years by L-Garde Inc. It has to be noted that the reflector antennas was mechanically oriented by the satellite itself acting through very long booms.

The link between two different satellites was developed by the Italian Space Agency (ASI) and NASA (Tethered Project) in year 1992 and 1996. The project envisaged the connection between two satellites via a 20 km tether cable.

In the present paper we will consider inflatable antenna "thetered" with a service satellite, but a different approach is adopted. The antenna itself is assumed to have within its own walls, notwithstanding their extremely small thickness, the possibility of generating forces, distributed on its surface, sufficient to rotate itself in the desired direction without any need of actions from the accompanying satellite.

## MATERIALS AND METHODS

### 1. The new antenna system

Let us imagine to metallize with a metallic coating just part of a sphere: in this way, a large reflecting surface will be realized. It would be even possible today to put in orbit balloons with a shape, for the coated portion, nearer to a parabola than to a sphere (Figure 1), as assumed in the studies concerning the QUASAT and the ARISE satellites [4].

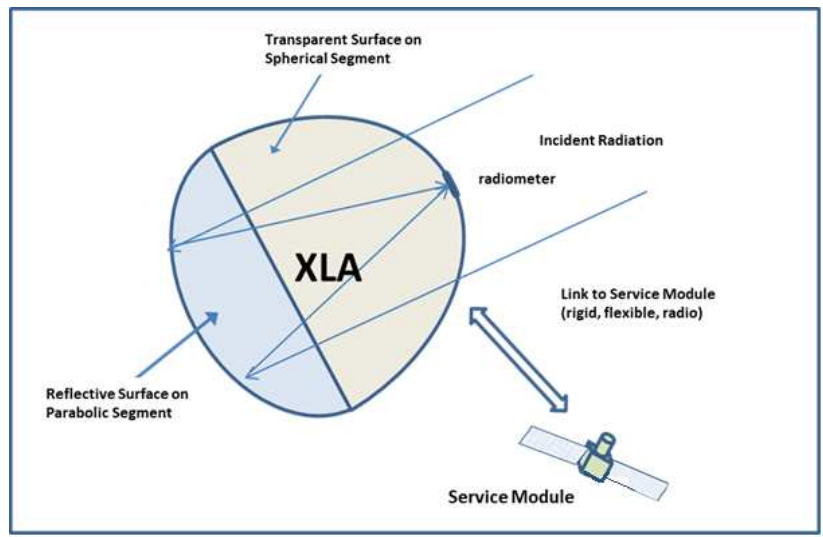

Figure 1. System configuration.

The antenna has the shape of a spheroid in which the southern hemisphere has a parabolic shape made of metallized fabric; the northern hemisphere, transparent to the electromagnetic radiation, has a radius of curvature such that the north pole coincides with the focus of the parabola.

However the main problem to be faced is how to orient the object itself if it has to behave like an antenna. Indeed the extreme smallness of the walls together with their extremely large dimensions imposes to exercise a diffuse, weak but regular force to slowly turn the antenna in the right direction in order not to introduce deformations and to induce instabilities in the envelope itself.

In a preliminary assessment, but sufficient for an evaluation of the proposal, is consider a simple spherical balloon in orbit at 2000 km, with a radius of the of 15 meters, above the earth's surface at the equator, where the earth's magnetic field $\vec{B}$ is approximately $1.375 \cdot 10^{-5}\ T$ [5].

In what follows a way of re-orienting the balloon in a chosen direction will be suggested although nothing will be said concerning the sensors to control the process: now the control system is not the main concern.

If, for the time being, we assume a 1 A electric current flowing in a conducting wire embedded in the balloon walls on a maximum radius circle, an electric coil is formed, interacting with the earth's magnetic field (Figure 2). The mechanic momentum $\vec{M}$ acting on the coil due to the presence of the magnetic field $\vec{B}$, is in turn equal to the product of the time derivative of the angular velocity, $\frac{d\vec{W}}{dt}$ (where $\vec{W}$ equals $\frac{d\theta}{dt}$ and $\theta$ is the angle between $\vec{B}$ and the plane of the ring), for the momentum of inertia of the body, $\vec{I}$, that is $\vec{M} = \vec{I} \times \frac{d\vec{W}}{dt}\ N\ m$

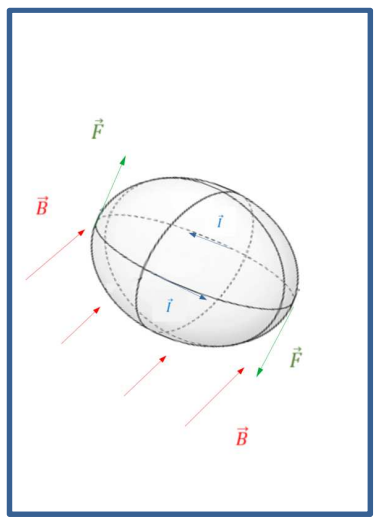

Figure 2. The forces acting in the presence

of a current $\vec{I}$ and the earth's magnetic field $\vec{B}$ are schematically indicated.

$\vec{M}$ equals the product of the current (1 A) for the area of the ring, $3.14 \cdot 15^2$ m², for the cosine of the angle $\theta$ between the earth's magnetic field and the plane identified by the ring itself, multiplied by $B = 1.375 \cdot 10^{-5} \, T$. For the time being let us assume to be interested only in evaluating the rotation for small angles between the magnetic field and the plane of the ring assumed co-planar at the beginning; in such a case, the cosine of the angle $\theta$ can be assumed constant and approximately equal to 1.

Assuming a mass of 50 kg for the entire balloon, mass entirely distributed for simplicity (only order of magnitudes are of interest at this point) along a ring of 15 m radius, $\vec{I}$ equals 11250 kg·m².

Consequently the above equation $\vec{M} = \vec{I} \times \frac{d\vec{W}}{dt}$ becomes:

$$M = 1 \cdot 3.14 \cdot 15^2 \cdot 1.375 \cdot 10^{-5} = 11250 \cdot \frac{dW}{dt} \, N \cdot m$$

Integrating twice,

$$\theta = 0.043 \cdot 10^{-5} \cdot t^2 \, rad$$

and if the rotation of interest equals $30^o$, putting $\theta = 30^o$ in the above equation and solving for t, it comes out that t equals approximately 18 minutes.

If we consider only the rotation equilibrium equation on the axis on the magnetic field plane and perpendicular to the field line, all perturbation torque in this context is neglected, we can write the following equation

$$I\ddot{\theta} = -i \sum B \sin\theta \, N \cdot m$$

where $\ddot{\theta}$ is the angular acceleration and I is the inertial torque [6]. This is a nonlinear second order differential equation that can be resolved only with numerical integration.

Assuming (Figure 3):

- On the ring is acting only the magnetic torque.
- At t = 0, the vector normal to the ring is orthogonal to the magnetic field vector.

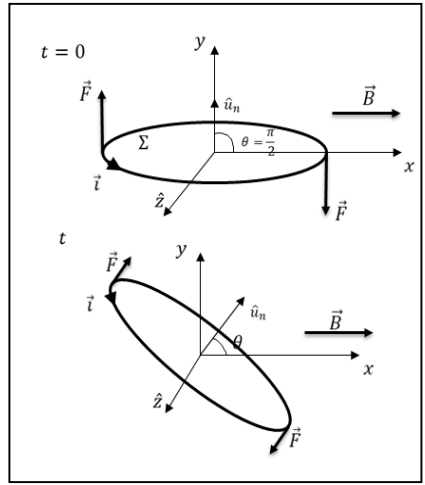

Figure 3. Magnetic torque.

We obtain with numerical integration that the total time to rotate the balloon of 50 deg is less than 2500 s (Figure 4).

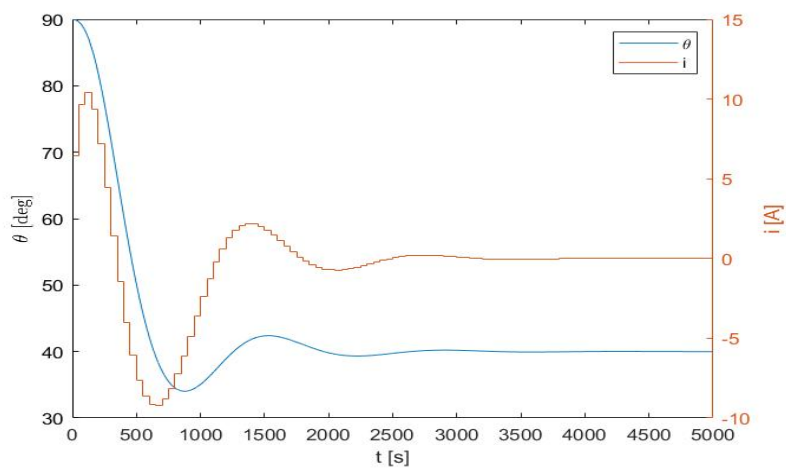

Figure 4. Numerical resolution of $I\ddot{\theta} = -i \sum B \sin\theta$

The above order of magnitude estimates show that significant rotations can be induced by the interaction of the earth's magnetic field with a current impressed in a conductor embedded in the balloon; in the computation presented, it has been assumed to excite a 1 A current but very different values of current, even significantly larger, may easily be excited, and, consequently, larger rotation velocity may be reached.

Let us now consider the energy required to move $30°$.

Supposing to orient the antenna of 30°, in a time of 10 minutes, in one of the 3 directions of space and supposing that the total kinetic energy is concentrated in a mass of 50 kg at 15 m from the center of rotation, the speed angle will be 0.0017 rad/sec, the relative speed is very low, equal to 2.6 cm/sec.

Considering that:

$$\begin{cases} l(t) = v_{0T} + \frac{1}{2}a_T t^2 \ m \\ v(t) = v_{0T} + a_T t \quad \frac{m}{s} \end{cases}$$

$$\begin{cases} \theta(t) = \omega_0 t + \frac{1}{2}\alpha t^2 \ rad \\ \omega(t) = \omega_0 + \alpha t \quad \frac{rad}{s} \end{cases}$$

if $\omega_0$ is 0 and θ(t) at time t= 0 it is 0

$$\omega(t) = 2 \cdot \frac{3.14}{6 \cdot 600} = 0.0017 \ \frac{rad}{s}$$

$$v(t) = 0.0017 \cdot 15 = 0.0255 \ \frac{m}{s}$$

And also

$$0.0255 \ \frac{m}{s} = 2.55 \ \frac{cm}{s}$$

The corresponding kinetic energy is 1.3 J , this quantity divided by 600 s, the time required to reach the 30° rotation, provides the average power required in watts significantly lower than the ohmic losses,

$$P = 1.645 \cdot 1^2 = 1.645 \ W$$

$$E_k = 0.0163 \ J$$

$$P = \frac{0.0163}{600} = 0.00027 \ W$$

These results indicate that orienting the antenna with the hypothesized long times implies very low rotation speeds and very modest energies.

The thickness of the walls combined with their extremely large surface makes it necessary to exert a diffused, weak but regular force to slowly orient the antenna in the right direction so as not to introduce deformations and induce instability in the casing itself.

## 2. Further problems and suggested solution

In order to orient the balloon in any desired direction, it should be necessary to dispose of 3 electric currents flowing in 3 rings at 90° one respect to the others. Of course this is possible without increasing too much the weight of the balloon (the total weight of 3 conductive wires (Cu, Ag, Au, etc.) with a 1 mm² cross section is approximately 6 kg). Furthermore, it will be necessary to activate the currents (and possibly modulate or even inverting them) in order to exactly orient the balloon in the desired direction; however, as we have stated, we are not considering these aspects now, requiring anyhow very simple electronic controls.

More important is to evaluate the energy needed to actuate the rotation. Let us assume to use 1 mm² Cu conductor. Each circle having a length of 94 m presents a total resistance of 1.6 W, consequently the power dissipated in heath to induce a 1 A current in the wire, is extremely small (less than 2 W).

Copper has a specific resistance of $0.017 \frac{\Omega \cdot mm^2}{m}$

A copper conductor with a length of one meter and a section of 1 mm² has an electrical resistance across its ends of 0.017 Ω. The resistance of a conductor is directly proportional to the length and inversely proportional to the section. The mathematical formula for calculating the electrical resistance of a generic conductor is:

$$R = \frac{\rho \cdot L}{s} \; \Omega$$

The coil length will be:

$$L = 2 \cdot 3.141 \cdot 15 = 94 \; m$$

Resistance of the copper wire of which the coil is made

$$R = \frac{0.017}{1} \cdot 94 = 1.645 \; \Omega$$

According to Ohm's law, power is:

$$P = 1.645 \cdot 1^2 = 1.645 \; W$$

The power dissipated by the wire will then be 1.645 W.

The mechanical energy to be provided for putting in rotation the balloon is easily evaluated. Let us assume that, at the end of the 30° rotation, the total kinetic energy is concentrated in a 50 kg mass at 15 m from the center of rotation having a velocity of $\frac{d\theta}{dt} \cdot 15 \; m/s$; therefore the final total kinetic energy is $0.5 \cdot 50 \cdot \left(\frac{d\theta}{dt} \cdot 15\right)^2$ where $\frac{d\theta}{dt}$, when $\theta = 30°$, equals $95 \cdot 10^{-5} \; rad/s$, as can be immediately deduced from the preceding computations; this quantity divided for the approximately 1000 s necessary to reach the 30° rotation, gives the needed average power, even significantly less than the ohmic losses.

In conclusion, the total electric power needed is not larger than a few watts for each ring to be activated, and can be easily provided by a few photovoltaic cells without significantly increasing the total weight of the balloon.

Another point to be addressed is the interaction of the metallic conductors with the walls of the balloon. As already stated, the interaction should be quite "delicate" in order not to deform the light structure of the balloon itself: the balloon should behave like a rigid body when put in rotation. Of course, this point should be carefully analyzed, however, some order of magnitude estimates can be performed. In order to put in rotation the structure, each 1 m of electric wire exercises a force f equal to $I \cdot B$. At the same time, the residual atmosphere within the balloon exerts a pressure p that in turn causes a force F at the interface between the balloon walls and the wire for each meter of the wire itself, of

$$F = 3.14 \cdot R^2 \cdot \frac{p}{2 \cdot 3.14 \cdot R} \quad N$$

f should be much smaller than F, consequently the pressure $i \cdot B \cdot \frac{2}{R} = 3.14 \cdot \frac{2}{R} = 0.18 \cdot 10^{-5}$ Pa should be much less than p.

Furthermore, in order to be a good electromagnetic reflector, the metallic coating should be thicker than the skin depth.

Finally, an elliptical orbit may be used (as for the proposed ARISE satellite) with a few thousands km perigee and a 40000 km apogee: the large angular orientation of the antenna/balloon may be implemented when the balloon is at the perigee where the earth magnetic field is bigger while small angular adjustments may be implemented around the apogee where the earth magnetic field is much smaller.

## RESULTS

The order of magnitude of data concerning the satellite, are proving the possibility of putting in orbit ultra-light extremely-large steerable antennas, having the mechanism of rotation embedded in their own walls, significantly simplifying the entire system. Furthermore, many of the functions needed on board an operative satellite may even be implemented on the balloons walls.

# DISCUSSION

The advantages of making inflatable satellites are reduced weight, the possibility of packaging for transport to space and low costs.

The satellite to be built is extremely light: less than 300 kg, very little compared to "traditional" satellites.

The satellite is easily steerable, unlike similar projects carried out in the past, using a low electric current generated by photovoltaic panels.

The satellite will be stationed in a low orbit about 2000 km above sea level, is passive, like a mirror, reflects radio waves from stations on earth and can be used for telecommunications, climate change studies and studies of sources that transmit by properties of their structures in microwave band.

# ACKNOWLEDGEMENTS

We thank C.I.F.S. (Consorzio Interuniversitario Fisica Spaziale) for their technical contribution.